  \documentclass[12]{article} %{cambridge6A}
  \usepackage{amsthm}
  \usepackage{graphicx}

\usepackage{amsfonts}

\begin{document}

  \title{Progress and Gravity: \\  Overcoming Divisions between General Relativity and Particle Physics and between Physics and HPS\footnote{Ch. 13, pp. 263-282 in Khalil Chamcham, Joseph Silk, John Barrow, and Simon Saunders, editors, \emph{The Philosophy of Cosmology}.  Cambridge University Press, Cambridge (2017); https://doi.org/10.1017/9781316535783.014.} }

  \author{J. Brian Pitts\\[3\baselineskip]
   Faculty of Philosophy, University of Cambridge}

\date{}

 % \frontmatter
  \maketitle

\pagebreak 

  \tableofcontents
%  \listoffigures
%  \listoftables
%  \listofcontributors

%%%%%%%%%%%%%%%%%%%%%%%%%%%%%%%%%%%%%%%%%%%%%%%%%%%%%%%%%%%%%%%%%%%%%%%%%%%%%%

\pagebreak

%%%%%%%%%%%%%%%%%%%%%%%%%%%%%%%%%%%%%%%%
\section{Introduction:  Science and the Philosophy of Science}

The ancient ``problem of the criterion''  is a chicken-or-the-egg problem regarding knowledge and criteria for knowledge, a problem that arises more specifically in relation to science and the philosophy of science.  How does one identify  reliable knowledge without a reliable method in hand? But how does one identify a reliable method without reliable examples of knowledge in hand?   Three possible responses to this  problem were entertained by Roderick Chisholm: one can be a  skeptic, or identify a reliable method(s) (``methodism''), or identify  reliable particular cases of knowledge (``particularism'')  \cite{ChisholmCriterion}.  But why should the best resources be all of the same type?  Mightn't some methods and some particular cases be far more secure than all  other methods and all other particular cases?  Must anything be completely certain anyway?  Why not mix and match, letting putative examples and methods tug at each other until one reaches (a personal?) reflective equilibrium?  

This problem arises for knowledge and epistemology, more specifically for science and the philosophy of science, and somewhere in between, for inductive inference.  
Reflective equilibrium  is  Nelson Goodman's method for induction (as expressed in John Rawls's terminology). One needn't agree with Goodman about deduction or take his treatment of induction to be both necessary and \emph{sufficient} to benefit from it.  He writes:
\begin{quote}  
%I have said that deductive inferences are justified by their conformity to valid general
%rules, and that general rules are justified by their conformity to valid inferences. But this circle is a virtuous one. 
%The point is that rules and particular inferences alike are justified by being brought into agreement with each other. 
\emph{A rule is amended if it yields an inference we are unwilling to accept; an inference is rejected if it violates a
rule we are unwilling to amend.} The process of justification is the delicate one of making mutual adjustments
between rules and accepted inferences; and in the agreement achieved lies the only justification needed for
either. 

All this applies equally well to induction. An inductive
inference, too, is justified by conformity to general rules,
and a general rule by conformity to accepted inductive
inferences. Predictions are justified if they conform to
valid canons of induction; and the canons are valid if they
accurately codify accepted inductive practice.
\cite[p. 64, emphasis in the original]{Goodman} 
\end{quote} 
Most scientists and (more surprisingly) even many philosophers  do not take Hume's problem of induction very seriously, though philosophers talk about it a lot.  As Colin Howson notes, philosophers often declare it to be insoluble and then proceed as though it were solved \cite{HowsonHume}.  I agree with Howson and Hans Reichenbach \cite[pp. 346, 347]{ReichenbachEP} that one should not let oneself off the hook so easily. That seems especially true in cosmology \cite{NortonInductionSpacetime}.  Whether harmonizing one's rules and examples is sufficient is less clear to me than it was to Goodman, but such reflective equilibrium surely is \emph{necessary}---though difficult and perhaps rare.  

My present purpose, however, is partly to apply Goodman-esque reasoning only to a \emph{special case} of the problem of the criterion, as well as to counsel unification within physical inquiry.  What is the relationship between  philosophy of science (not epistemology in general) on the one hand, and scientific cosmology and its associated fundamental physics, especially gravitation and space-time theory (not knowledge in general) on the other?  Neither dictation from philosopher-kings to scientists (the analogue of methodism) nor complete deference to scientists by philosophers (the analogue of particularism) is Goodman's method.  It is not popular for philosophy to give orders to science, but it once was. The reverse is more fashionable, a form of scientism or at least a variety of naturalism.  I hope to show by examples how sometimes each side should learn from the other.  

%Since the early 20th century decline of Kantianism in the wake of Moritz Schlick's critique from General Relativity (on which more below), 

While Goodman's philosophy has a free-wheeling relativist feel that might make many scientists and philosophers of science nervous, one finds similar views expressed by a law-and-order philosopher of scientific progress, Imre Lakatos.  According to him, we should seek 
\begin{quote} a pluralistic system of authority, partly because the wisdom of the scientific jury and its case law has not been, and cannot be, fully articulated by the philosopher's statute law, and partly because the philosopher's statute law may occasionally be right when the scientists' judgment fails.  \cite[p. 121]{LakatosHistory} \end{quote}  Thus there seems to be no irresistible pull toward relativism in seeking reflective equilibrium rather than picking one side always to win automatically.

%%%%%%%%%%%%%%%%%%%%%%%%%%%%%%%%%%%%%%%%%%%%%%%%%%%%%%%%%%%%%%%%%%%%

\section{Healing the GR \emph{vs.} Particle Physics Split}

A second division that should be overcome to facilitate the progress of knowledge about gravitation and space-time is the general relativist \emph{vs.} particle physicist split.  Carlo Rovelli discusses 
\begin{quote} \ldots the different understanding of the world that the
particle physics community on the one hand and the relativity community on
the other hand, have. The two communities have made repeated and sincere
efforts to talk to each other and understanding each other. But the divide
remains, and, with the divide, the feeling, on both sides, that the other side is
incapable of appreciating something basic and essential\ldots. 
 %Hopefully, the recent successes of both lines will force the two sides, finally, to face the problems that the other side considers prioritary\ldots.  
\cite{Rovelli}  \end{quote}

This split has a fairly long history going back to Einstein's withdrawing from mainstream fundamental physics from the 1920s---that largely being quantum mechanics, relativistic quantum mechanics and quantum field theory.  A further issue  pertains to the gulf between how Einstein actually found his field equations (as uncovered by recent historical work \cite{Renn,RennSauer,RennSauerPathways})  and the much better known story that Einstein told retrospectively. Work by J\"{u}rgen Renn \emph{et al.} has recovered the importance of Einstein's ``physical strategy'' involving a Newtonian limit, an analogy to electromagnetism, and a quest for energy-momentum conservation; this strategy ran along side the better advertised  mathematical strategy  emphasizing his principles (generalized relativity, general covariance, equivalence, \emph{etc.}).  Einstein's reconstruction of his own past is at least in part a persuasive device in defense of his somewhat lonely quest for unified field theories \cite{vanDongenBook}.  Readers with an eye for particle physics will not miss the similarity to the later successful derivations of Einstein's equations as the field equations of a massless spin $2$ field assumed initially to live in flat Minkowski space-time \cite{Feynman}, in which the resulting dynamics merges  the gravitational potentials with the flat space-time geometry such  that only an effective curved geometry appears in the Euler-Lagrange equations.  
One rogue general relativist has recently opined:
\begin{quote} 
HOW MUCH OF AN ADVANTAGE did Einstein gain over his colleagues by his mistakes?  Typically, about ten or twenty years. For instance,
if Einstein had not introduced the mistaken Principle of Equivalence and
approached the theory of general relativity via this twisted path, other physicists would have discovered the theory of general relativity some twenty
years later, via a path originating in relativistic quantum mechanics. \cite[p. 334, capitalization in the original]{OhanianEinsteinMistakes}. \end{quote}

It is much clearer that these derivations work to give Einstein's equation than it is what they mean.  Do they imply that one needn't and perhaps shouldn't ever have given up flat space-time?  Do they, on the contrary, show that theories of gravity in flat space-time could not succeed, because their best effort turns out to give curved space-time after all \cite{EhlersMehra}?  Such an argument is clearly incomplete without contemplation of massive spin 2 gravity \cite{OP,FMS}.  But it might be persuasive if massive spin 2 gravity failed---as it seemed to do roughly when Ehlers wrote (not that he seems to have been watching).  But since 2010 massive spin-$2$ gravity seems potentially viable again \cite{deRhamGabadadze,HassanRosenNonlinear,MaheshwariIdentity} (though some new issues exist).  Do the spin-$2$ derivations of Einstein's equations  suggest a conventionalist view that there is no fact of the matter about the true geometry \cite[[pp. 112, 113]{Feynman}?  Much of one's assessment of  conventionalism will depend on what one takes the modal scope of the discussion to be:  should one consider only one's best theory (hence the question is largely a matter of exegeting General Relativity, which will favor curved space-time), or should one consider a variety of theories?  According to John Norton, the philosophy of geometry is not an enterprise rightly devoted to giving  a spurious air of necessity to whatever theory is presently our best \cite[pp. 848, 849]{Norton}.  Such a view suggests the value of a broader modal scope for the discussion than just our best current theory.  On the other hand, the claim has been made that the transition from Special Relativity to General Relativity is as unlikely to be reversed as the transition from classical to quantum mechanics \cite[pp. 84, 85]{EhlersMehra}.  If one aspires to proportion belief to evidence, that is a startling claim.  The transition from classical to quantum mechanics was motivated by grave empirical problems; there now exist theorems (no local hidden variables) showing how far any empirically adequate physics \emph{must} diverge from classical. But a constructive derivation of Einstein's equations from a massless spin-$2$ shows that one can naturally recover the phenomena of GR without giving up a special relativistic framework in a sense.  The cases differ as twilight and day.  Ehlers's remarks are useful, however, in alerting one for Hegelian undercurrents or other doctrines of inevitable progress in the general relativity literature. A classic study of doctrines of progress is (\cite{BuryProgress}).  
  %The Locke-Hume aspiration to proportion belief to evidence seems more scientifically judicious, however.  

%Bury progress, irreversible progress EhlersMehra
%
%
%One can likely hear echoes of early 20th century French and German political struggles.  In those contexts conventionalism geometry was both one of the best ideas in the philosophy of science  and a potential resource for traditionalists to preserve their favorite doctrines from scientific refutation.  Advocates of progressive causes (including the logical empiricists Moritz Schlick and Hans Reichenbach) would thus feel pressure to mold their views of science in ways that would favor their own causes and/or deprive traditionalists of intellectual resources.  Fortunately such political struggles were resolved.  Open-minded evidence-based theory choice, as widely endorsed in the philosophy of science, now can take place.  

%%%%%%%%%%%%%%%%%%%%%%%%%%%

\section{Bayesianism, Simplicity, and Scalar \emph{vs.} Tensor Gravity}

While Bayesianism has made considerable inroads in the sciences lately, it is helpful to provide a brief sketch before casting further discussion in such terms.  I will sketch a rather simple version---one that might well be inadequate for science, in which one sometimes wants uniform probabilities over infinite intervals and hence might want infinitesimals, for example.   Abner Shimony's tempered personalism discusses useful features for a scientifically usable form of Bayesianism, including open-mindedness (avoiding prior probabilities so close to $0$ or $1$ that evidence cannot realistically make much difference) \cite{ShimonyInference}) and assigning non-negligible prior probabilities to seriously proposed hypotheses.

With such qualifications in mind, one can proceed to the sketch of Bayesianism.  One isn't equally sure of everything that one believes, so why not have degrees of belief, and make them be real numbers between $0$ and $1$?  Thus one can hope to mathematize logic in shades of gray \emph{via} the  probability calculus.  Bayes's theorem  can be applied to a theory $T$ and evidence $E$:
\begin{equation} P(T|E) = P(T) \frac{ P(E|T) }{ P(E) }.  
\end{equation}
One wakes up with degrees of belief in all theories (!), ``prior probabilities.''  One opens one's eyes, beholds evidence $E$, and goes to bed again. While asleep one revises degrees of belief from priors $P(T)$ to posterior probabilities $P(T|E)$.  Today's $P(T|E)$ becomes tomorrow's prior $P(T)^{\prime}$. Then one does the same thing tomorrow, getting some new evidence $E^{\prime},$ \emph{etc.}  
Now the priors $P(T)$ might be partly subjective.   If there are no empirically equivalent theories and everyone is open-minded, then eventually evidence should bring convergence of opinion over time (though maybe not soon).  

A further wrinkle in the relation between evidence and theory comes from looking at the denominator of Bayes's theorem,  $P(E)= P(E|T) P(T) + P(E|T_1) P(T_1) + P(E|T_2) P(T_2) +\ldots.$  While one might have hoped to evaluate evidence theory $T$ simply in light of evidence $E$, this expansion of $P(E)$ shows that such an evaluation is typically undefined, because one must spread degree of belief $1-P(T)$ among the competitors $T_1,$ $T_2$, \emph{etc.}  Hence the predictive likelihoods $P(E|T_1)$, \emph{etc.},  subjectively weighted, appear unbidden in the test of $T$ by $E$.  Theory testing generically is \emph{comparative}, making essential reference to rival theories. This fact is sometimes recognized in scientific practice, but Bayesianism can alert one to attend to the question more systematically. %That fact can be especially interesting if the competitors aren't all equally objective.  

%%%%%%%%%%%%%%%%%%%%%%%%%%%%%%%%%%%%%%%%%%%%%%%%%%%%%%%%%%%%%%%%%%%%%%%%%%%%%%%%%%%%%%%%%%%%

 Scientists and philosophers tend to like simplicity. Simplicity might not be objective, but there is significant agreement regarding  scientific examples.  That is a good thing, because there are lots of theories, especially lots of complicated ones, way too many to handle.  If degrees of belief are real numbers (not infinitesimals), then normalization $\Sigma_i P_i=1$ requires lots of $0$'s and or getting ever closer to $0$ on some ordering \cite[pp. 209, 210]{EarmanBayes}.  There is no  clear reason for prior plausibility to peak away from the simple end. Plausibly, other things equal, simpler theories are more plausible \emph{a priori}, getting a  higher  prior $P(T)$ in a Bayesian context. Such considerations are vague, but the alternatives are even less principled.  

                    %              Swinburne, simplicity as evidence for truth?

%%%%%%%%%%%%%%%%%%%%%%%%%%%%%%%%%%%%%%%%%%%%%%%%%%%%%%%%%%%%%%%%%%%%%%%%%%%%%%%%%%%%%%%%%%%%%%%%%%%%

One can now apply Bayesian considerations to gravitational theory choice in the 1910s.  One recalls that Einstein had some arguments against a scalar theory of gravity, which motivated his generalization to a tensor theory.  Unfortunately they don't work.  As Domenico Giulini has said, 
 \begin{quote}  
On his way to General Relativity, Einstein gave several arguments as to why a special-relativistic
theory of gravity based on a massless scalar field could be ruled out merely on grounds of theoretical
considerations. We re-investigate his two main arguments, which relate to energy conservation and
some form of the principle of the universality of free fall. We find such a theory-based \emph{a priori}
abandonment not to be justified. Rather, the theory seems formally perfectly viable, though in clear
contradiction with (later) experiments.  \cite[emphasis in original]{GiuliniScalar}  \end{quote}

Scalar (spin-$0$) gravity is simpler than rank-2 tensor (spin-$2$). Having one potential is simpler than having $10$, especially if they are self-interacting. With Einstein's help, Gunnar Nordstr\"{o}m eventually proposed a scalar theory that avoided the theoretical problems mentioned by Giulini.  Given simplicity considerations, Nordstr\"{o}m's theory was more probable than Einstein's \emph{a priori}: $P(T_N)>P(T_{GR}).$  Einstein's  further criticisms are generally matters of taste.  So prior to evidence for General Relativity, it was more reasonable to favor Nordstr\"{o}m's theory.  As it actually happened, Einstein's `final' theory and the evidence from Mercury both appeared in November 1915, leaving little time for this logical moment in actual history. Einstein's earlier \emph{Entwurf} theory \cite{EinsteinEntwurf} could be faulted for having negative-energy degrees of freedom and hence likely being unstable (a problem with roots in Lagrange and Dirichlet \cite{Morrison}), though apparently no one did so.

Where was the progress of scientific knowledge---truth held for good reasons?
Mercury's perihelion gave non-coercive evidence confirming  GR and disconfirming Nordstr\"{o}m's theory. It was possible to save Nordstr\"{o}m's theory using something like dark matter, matter (even if not dark---Seeliger's zodiacal light) of which the mass had been neglected \cite{RoseveareMercury}.  Hence there was scope for rational disagreement because Nordstr\"{o}m's theory was antecedently more plausible $$P(T_N)>P(T_{GR})$$ but evidence favored Einstein's non-coercively $$0<P(E_{Merc}|T_N) < P(E_{Merc}|T_{GR}).$$    
The scene changed in 1919 with the bending of light, which falsified Nordstr\"{o}m's theory:  $P(E_L|T_N) = 0.$  There were not then other plausible theories that predicted light bending, so $P(E_L|T_{GR})\approx 1 >> P(E_L)$.  It is possible to exaggerate the significance of this result, as happened popularly but perhaps less so academically \cite{BrushLightBending}, where a search for plausible rival theories that also predicted light bending was made. 
(Bertrand Russell may have considered Whitehead's to be an example \cite[pp. 75-80]{RussellMatter}.)  
Unfortunately many  authors wrongly take Einstein's arguments against scalar gravity seriously \cite{GiuliniScalar}.  In the long run one does not make  \emph{reliable rational} progress by siding with genius as soon as possible: Einstein made many mistakes (often correcting them himself), some of them lucky \cite{OhanianEinsteinMistakes} (such as early rejection of scalar theories), followed by barren decades.  
 Given this Bayesian sketch, it was rational to prefer GR over Nordstr\"{o}m's scalar theory only when evidence from Mercury was taken into account, and not necessarily even then.  The bending of light excluded scalar theories but did not exclude possible rival tensor theories.

%%%%%%%%%%%%%%%%%%%%%%%%%%%%%%%%%%%%%%%%%%%%%%%%%%%%%%%%%%%%%%%%%%%%%%%%%%%%%%%%%%%%%%%%%%%%%%%%%%%%%%%%%%%%%%%%%%%%%%%%%%%%%%

\section{General Relativity Makes Sense about Energy}

 Resolving conceptual problems is  a key part of scientific progress \cite{LaudanProgress}.
In the 1910s and again in the 1950s controversy arose over the status of energy-momentum conservation laws of General Relativity.  Given Einstein's frequent invocation of energy-momentum conservation in his process of discovery leading to General Relativity  \cite{EinsteinEntwurf,RennSauer,BradingConserve,RennSauerPathways}, as well as his retrospective satisfaction  \cite{EinsteinHamiltonGerman}, this is ironic.  Partly in response to Felix Klein's dissatisfaction, Emmy Noether's theorems appeared \cite{Noether}.  Her first theorem says that a rigid symmetry yields a continuity equation. Her second says that a wiggly symmetry yields an identity among Euler-Lagrange equations, making them not all independent.  For General Relativity there are $4$ wiggly symmetries, yielding the contracted  Bianchi identities $\nabla_{\mu} \mathcal{G}^{\mu}_{\nu} \equiv 0.$  In the wake of the conservation law controversies there emerged the widespread view that gravitational energy exists, but it `is not localized'.  This phrase appears to mean that gravitational energy isn't anywhere in particular, though descriptions of it often do have locations.  That puzzling conclusion is motivated by mathematical results suggesting that where gravitational energy is depends on an arbitrary conventional choice (a coordinate system), and other results that the total energy/mass does not.

While the energy nonlocalization lore is harmless enough as long as one knows the mathematical results on which it is based, it has self-toxifying quality.  Having accepted that gravitational energy isn't localized, one is likely to look askance at the Noether-theoretic calculations that yield it:  pseudotensors.  The next generation of textbooks might then dispense with the calculations while retaining the lore verbally. Because the purely verbal lore is mystifying, at that point one formally gives license to a variety of doubtful  conclusions.  Among these are that because General Relativity lacks conservation laws, it is false---a claim at the origins of the just-deceased  Soviet/Russian academician A. A. Logunov's high-profile dissent  \cite{LogunovEnergy1977}.  
One also hears (for references see (\cite{EnergyGravity}) that the expansion of the universe, by virtue of violating conservation laws, is false (a special case of Logunov's claim).  One hears that the expansion of the universe is a resource for creation science by giving a providing a heat sink for energy from rapid nuclear decay during Noah's Flood.  Finally, one hears that General Relativity is more open to the soul's action on the body than is earlier physics, because the soul's action violates energy conservation, but General Relativity already discards energy conservation anyway.  That last claim is almost backwards, because Einstein's equations are logically equivalent to energy-momentum  conservation laws \cite{Anderson}.  (If one wants souls to act on bodies, souls had better couple to gravity also.)  The question whether vanishing total energy of the universe (given certain topologies) would permit it to pop into being spontaneously is also implicated.

Given that Noether's theorems---the first, not just the second---apply to GR, can one interpret the continuity equations sensibly and block the unfortunate inferences?   The Noether operator generalizes canonical stress-energy tensor to give conserved quantities due to symmetry vector fields $\xi^{\mu}$  \cite{BergmannConservation,TrautmanConserve,SorkinStress,GoldbergReview,SzabadosSparlingHAS}.  
For simpler theories than GR, the Noether operator is a weight $1$ tangent vector density $\mathfrak{T}^{\mu}_{\nu} \xi^{\nu}$, so the divergence of the current $\partial_{\mu} (\mathfrak{T}^{\mu}_{\nu} \xi^{\nu})$ is tensorial (equivalent in all coordinate systems), and, for symmetries $\xi^{\nu},$ there is conservation $\partial_{\mu} (\mathfrak{T}^{\mu}_{\nu} \xi^{\nu})=0$.  GR (the Lagrangian density, not the metric!) has \emph{uncountably many} `rigid' translation symmetries  $x^{\mu} \rightarrow x^{\mu} + c^{\mu}$, where $c^{\mu},_{\nu}=0,$ for any coordinate system, preserving the action $S=\int d^4x \mathcal{L}$. 
These uncountably many symmetries yield uncountable conserved energy-momentum currents.  Why can't they all be real?  The lore holds that because there are infinitely many currents, really there aren't any.  But just because it's infinite doesn't mean it's 0 (to recall an old phrase). Getting $\infty=0$ requires an extra premise, to be uncovered shortly.  For GR, the Noether operator is a conserved but \emph{nontensorial differential} operator on $\xi$, depending on $\partial \xi$ also.   Hence one obtains coordinate-dependent results, with energy density vanishing at an arbitrary point, \emph{etc.}, the usual supposed vices of  pseudotensors.  If one expects only one energy-momentum (or rather, four), it should be tensorial, with the transformation law relating faces in different coordinates. But Noether tells us that there are \emph{uncountably many} rigid translation symmetries.

If one simply `takes Noether's theorem literally' \cite{EnergyGravity} (apparently novelly, though Einstein and Tolman \cite{TolmanEnergy} said  nice things about pseudotensors), then  uncountably many symmetries imply uncountably many conserved quantities.  How does one get $\infty=0$?  By assuming that the infinity of conserved energies are all supposed to be faces of the same conserved entity with a handful of components---the key tacit premise of uniqueness.  Suppose that one is told in Tenerife that  ``George is healthy'' \&  ``Jorge est\'{a} enfermo'' (is sick). If one expects the two sentences to be equivalent under translation (analogous to a coordinate transformation), then one faces a contradiction:  George is healthy and unhealthy.  But if  George and Jorge then walk into the room together, there is no tension: $George \neq Jorge.$  An expectation of uniqueness underlies most objections to pseudotensors, but it is unclear what justifies that expectation.  Making more sense of energy conservation makes its appearance in Einstein's physical strategy in finding his  field equations less ironic. Indeed, conservation  due to gauge invariance is a key step in spin-$2$ derivations, which improve on Einstein's physical strategy  \cite{EinsteinEntwurf,Deser,SliBimGRG}.   Noether commented on \emph{converses} to her theorems \cite{Noether}; one should be able to derive Einstein's equations from the conservation laws, much as the spin-$2$ derivations do using symmetric gravitational stress-energy (hence perhaps needing Belinfante-Rosenfeld technology).

But what is  the point of believing in gravitational energy unless it does energetic things?  Can it heat up a cup of coffee?  Where is the physical interaction?\footnote{This question was articulated Erik Curiel.}   
Fortunately these questions have decent answers:  gravitational energy is roughly the nonlinearity of Einstein's equations, so it mediates the gravitational self-interaction.

Why did Hermann Bondi changed from a skeptic to a believer in energy-carrying gravitational waves \cite{BondiNature}?\footnote{I thank Carlo Rovelli for mentioning Bondi.}  Given a novel plane wave solution of Einstein's equations in vacuum, his equation (2), he wrote:
\begin{quote} there is a non-flat region of space between two flat
ones, that is, we have a plane-wave zone of finite
extent in a non-singular metric satisfying Lichnerowicz's
criteria [reference suppressed].
Consider now a set of test particles at rest in metric
(2) before the arrival of the wave. \cite{BondiNature}
\end{quote} 
After the passage of the wave, there is relative motion.
\begin{quote}
Clearly, this system of test particles in relative motion contains energy that could be used,
for example, by letting them rub against a rigid friction disk carried by one of them. \cite{BondiNature}
\end{quote}  
This has argument carried the day with most people since that time:  gravitational energy-transporting waves exist and do energetic things.

This argument has roots in Feynman  \cite{WikipediaStickyBead} \cite[p. 143]{DeWitt57ChapelHill} \cite[xxv, xxvi]{Feynman} \cite{KennefickWaves}. John Preskill and Kip Thorne, drawing partly on unpublished sources, elaborate. 
\begin{quote}
At Chapel Hill, Feynman addressed this issue in a pragmatic way, describing how a gravitational wave antenna could in principle be designed that would absorb the energy ``carried'' by the wave [DeWi 57, Feyn 57].  In Lecture 16, he is clearly leading up to a description of a variant of this device, when the notes abruptly end:  ``We shall therefore show that they can indeed heat up a wall, so there is no question as to their energy content.''  A variant of Feynman's antenna was published by Bondi [Bond 57] shortly after Chapel Hill (ironically, as Bondi had once been skeptical about the reality of gravitational waves), but Feynman never published anything about it.   The best surviving description of this work is in a letter to Victor Weisskopf completed in February, 1961 [Feyn 61].  %This letter contains some of the same material as Lecture 16, but then goes a bit further, and derives the formula for the power radiated in the quadrupole approximation (a result also quoted at Chapel Hill).  Then the letter describes Feynman's gravitational wave detector:  It is simply two beads sliding freely (but with a small amount of friction) on a rigid rod.  As the wave passes over the rod, atomic forces hold the length of the rod fixed, but the proper distance between the two beads oscillates.  Thus, the beads rub against the rod, dissipating heat.  Feynman included the letter to Weisskopf in the material he distributed to the students taking the course. 
\cite[xxv, xxvi]{Feynman}  
\end{quote}

Gravitational energy in waves exists in GR, and one of the main objections to localization can be managed by taking Noether's theorem seriously:  there are infinitely many symmetries and energies. Another problem is the non-uniqueness of the pseudotensor, which one might address with either a best candidate (as in Joseph Katz's work) or a physical meaning for the diversity of them in relation to boundary conditions (James Nester \emph{et al.}). Even scalar fields have an analogous problem \cite{ImprovedEnergy}.
 With hope there as well, energy in GR, though still in need of investigation, isn't clearly a serious conceptual problem anymore. That is scientific progress \`{a} la Laudan.

%
%The 1957 Chapel Hill conference report informs us:
%\begin{quote}
%The question of the absorption and production of gravitational waves was
%raised again. FEYNMAN discussed a device which would absorb gravitational energy,
%provided one assumes the existence of gravitational radiation (but as he pointed
%out, ``My instincts are that if you can feel it, you can make it.''). For this purpose one can use a result already presented
%at an earlier session that the displacement $\eta$ of a particle in the path of a gravitational wave satisfies the differential
%equation $$\frac{d^2 \eta}{dt^2} = R^{\,\,\,a}_{0ba} \eta \,\,. $$ 
%A particle situated initially near a long light rod, oriented parallel to the propagation direction, could be made to scrape against the rod by the transverse-transverse
%wave amplitudes. \cite[p. 143]{DeWitt57ChapelHill} % report says not to quote without permission from author, which would be (now) Cecile DeWitt
%\end{quote}
%

%
%%%%%%%%%%%%%%%%%%%%%%%%%%%%%%%%%%%%%%%%%%%%%%%%%%%%%%

\section{Change in Hamiltonian General Relativity}

Supposedly, change is missing in Hamiltonian General Relativity   \cite{EarmanMcTaggart}. That seems problematic for two reasons:  change is evident in the world, and change is evident in Lagrangian GR in that most solutions of Einstein's equations lack a time-like Killing vector field 
  \cite[p. 352]{Ohanian}.   A conceptual problem straddling the internal \emph{vs.} external categories is ``empirical incoherence,'' being self-undermining.
  According to Richard Healey, 
  \begin{quote} [t]here can be no reason whatever to accept any theory of gravity...which entails that there can be no observers, or that observers can have no experiences,  some occurring later than others, or that there can be no change in the mental state of observers, or that observers cannot perform different acts at different times.   It follows that there can be no reason to accept any theory of gravity ...which entails that there is no time, or no change. \cite[p. 300]{HealeyGRchangelessincoherent} \end{quote}
Hence accepting the no-change conclusion about Hamiltonian GR would undermine reasons to accept Hamiltonian GR.  Change in the world is safe.  But what about the surprising failure of Hamiltonian-Lagrangian equivalence? 

A key issue involves where one looks for change, and relatedly, one what means by ``observables.''  According to Earman (who would not dispute the point about the scarcity of solutions with time-like Killing vectors),  
``[n]o genuine physical magnitude countenanced in GTR changes
over time.'' \cite{EarmanMcTaggart} Since the lack of time-like Killing vectors implies that the metric does change, clearly genuine physical magnitudes must be scarce, rarer than tensors.  Tim Maudlin appeals to change in solutions to Einstein's equations: ``stars collapse, perihelions precess, binary star systems radiate
gravitational waves\ldots'' but ``a sprinkling of the magic powder of the
constrained Hamiltonian formalism has been employed to resurrect the decomposing flesh of McTaggart.\ldots'' \cite{MaudlinMcTaggart} 
Maudlin's appeal to common sense and Einstein's equations is  helpful, as is Karel Kucha\v{r}'s \cite{KucharCanonical93}, but one needs more detail, motivation and (in light of Kucha\v{r}'s disparate treatments of time and space) consistency.

Fortunately the physics  reveals a relevant controversy, with reformers recovering  Hamiltonian-Lagrangian equivalence  \cite{MukundaGaugeGenerator,CastellaniGaugeGenerator,SuganoGaugeGenerator,GraciaPons,PonsSalisburyShepley,PonsReduce,PonsSalisbury,PonsSalisburySundermeyerFolklore}. Hamiltonian-Lagrangian equivalence was manifest originally  \cite{RosenfeldQG,AndersonBergmann,SalisburyRosenfeldMed}; its loss needs study. In constrained Hamiltonian theories \cite{Sundermeyer}, some canonical momenta are (in simpler cases) just $0$  due to independence of $\mathcal{L}$ from some $\dot{q}_i$; these are ``primary constraints.''  In many cases of interest (including electromagnetism, Yang-Mills fields, and General Relativity), some functions of $p,$ $q$, $\partial_i p,$ $\partial_i q,$ $\partial_j \partial_i q$ are also $0$ in order to preserve the primary constraints over time. Often these `secondary' (or higher) constraints are familiar, such as the phase space analog $\partial_i p^i =0$ of Gauss's law $\nabla \cdot \vec{E}=0,$ Gauss-Codazzi equations embedding space into space-time in General Relativity, \emph{etc.}
Some constraints have something to do with gauge freedom (time-dependent redescriptions leaving the state or history alone). One takes Poisson brackets ($q$, $p$ derivatives) of all constraints pairwise. If the result is in every case $0$ (perhaps using the constraints themselves), then all constraints are ``first-class,'' as in Clerk Maxwell's electromagnetism, Yang-Mills, and GR in their most common formulations.  In General Relativity, the Hamiltonian, which determines time evolution, is nothing but a sum of first-class constraints (and boundary terms).
 Given that first-class constraints are related to gauge transformations, the key question is how they are related.  Does each do so by itself, or do they rather work as a team? There is a widespread belief that each does so individually  \cite{DiracLQM}. Then the Hamiltonian generates a sum of redescriptions leaving everything as it was, hence there is no real change.  This is a classical aspect of the `problem of time'.   
  Some try  to accept this conclusion, but recall Healey's critique.  

Because Einstein's equations and common sense agree on real change, something must have gone wrong in Hamiltonian GR or the common interpretive glosses thereon, but what?  Here the Lagrangian-equivalent reforming party has given most of the answer, namely, that what generates gauge transformations is not each first class constraint separately, but the gauge generator $G$, a  specially tuned sum of first-class constraints, secondary \emph{and primary} \cite{AndersonBergmann,CastellaniGaugeGenerator,PonsSalisburyShepley,PonsDirac,PonsSalisburySundermeyerFolklore}.  Thus electromagnetism has two constraints at each point but only one arbitrary function; GR has eight constraints at each point but only four arbitrary functions.  Indeed one can show that an isolated first-class constraint makes a mess \cite{FirstClassNotGaugeEM,GRChangeNoKilling}, such as spoiling the relation expected relation $\dot{q}=\frac{\delta H}{\delta p}$ making the canonical momentum equal to the electric field or the extrinsic curvature of space within space-time.  These canonical momenta  are  auxiliary fields in the canonical action $\int dt d^3x(p \dot{q}-\mathcal{H}),$ and hence get their physical meaning from $\dot{q}.$
Because  each first-class constraint makes a physical difference by itself (albeit a bad one), the GR Hamiltonian no longer is forced to be generate a gauge transformation by being a sum of them.  There is change in the Hamiltonian formalism  whenever there is no time-like Killing vector, just as one would expect from Lagrangian equivalence. 
\begin{quote}  We have been guided by the principle that the Lagrangian and Hamiltonian formalisms should be equivalent\ldots in coming to the conclusion that they in fact are.  \cite[p. 17]{PonsReduce}  \end{quote} 
By the same token, separate first-class constraints don't change  $p \dot{q}-\mathcal{H}$ by (at most)  a total derivative, but $G$  \emph{does} \cite{FirstClassNotGaugeEM,GRChangeNoKilling}. 
 
To get changing observables in GR, one should recall the distinction between internal and external symmetries.  Requiring that observables have $0$ Poisson bracket with the electromagnetic gauge symmetry generator is just to say that things that we cannot observe (in the ordinary sense) are unobservable (in the technical sense).  By contrast, requiring that observables have $0$ Poisson bracket with the gauge generator in GR implies that the Lie derivative of an observable is $0$ in every direction.  Thus anything that varies spatiotemporally is ``unobservable''--a result that cannot be taken seriously. The problem is generated by hastily generalizing the definition from internal to external symmetries. Instead one should permit observables to have Lie derivatives that are not $0$ but just the Lie derivative of a geometric object---an infinitesimal Hamiltonian form of the identification of observables with geometric objects in the classical sense \cite{Nijenhuis}, \emph{viz.},  set of components in each coordinate system and a transformation law.

%%%%%%%%%%%%%%%%%%%%%%%%%%%%%%%%%%%%%%%%%%%%%%%%%%%%%%%%%%%%%%%%%%%%%%%%%%%%%%%%%%%%%%%%%%%%%%%%%%%%%%%%%%%%%%%%%%%%%%%%%%%%%%%%%%%%%%%%%%%%%%%%%%%%%%%%%%%%%%%%%%%

\section{Einstein's Real $\Lambda$ Blunder in 1917} 

One tends to regard perturbative expansions and geometry as unrelated at best, if not  negatively related. 
\begin{quote}
The advent of supergravity [footnote suppressed] %\footnote{Supergravity is a family of theories including gravity with a rather ambitious symmetry that relates every boson to a fermion (of equal mass!---at least at first glance) and \emph{vice versa}.  }
 made relativists and particle physicists meet.  For many this was quite a new experience since very different languages were used in the two communities.  Only Stanley Deser was part of both camps.  The particle physicists had been brought up to consider perturbation series while relativists usually ignored such issues.  They knew all about geometry instead, a subject particle physicists knew very little about.  \cite[p. 40] {BrinkDeserSupergravity}  \end{quote}
  But some examples will show how perturbative expansions can help to reveal the geometric content of a theory that is otherwise often misunderstood, can facilitate the conception of novel geometric objects that one might otherwise fail to conceive, and permit conceptual and ontological insight.

Perturbative expansions can help to reveal the geometric content of a theory that one might well miss otherwise. Einstein in his 1917 cosmological constant paper first reinvented a long-range modification of Newtonian gravity \cite{EinsteinCosmological}---one might call it (anachronistically) nonrelativistic massive scalar gravity---previously proposed in the 19th century by Hugo von Seeliger and Carl Neumann.  But he then made a false analogy to his new cosmological constant $\Lambda$, a mistake never detected till the 1940s \cite{Heckmann}, not widely discussed till the 1960s, and still committed at times today.  According to Einstein,  $\Lambda$ was ``completely analogous to the extension of the Poisson equation to $ \Delta \phi - \lambda \phi = 4\pi K \rho$ '' \cite{EinsteinCosmological}.
Engelbert Sch\"{u}cking, a former student of Heckmann, provided a firm evaluation.  
``This remark was the opening line in a bizarre comedy of errors.'' \cite{Schucking}
The problem is that  $\Lambda$ is predominantly $0th$ order in $\phi$ (having a leading constant term), whereas the modified Poisson is $1$st order in $\phi.$
 $\Lambda$ gives a weird quadratic potential for a point source, but the modified Poisson equation gives a massive graviton with plausible Neumann-Yukawa exponential fall-off \cite{FMS,Schucking}.  ``However generations of physicists have parroted this nonsense.'' \cite{Schucking}
Massive theories of gravity generically involve 2 metrics, whereas $\Lambda$ involves only one. Understanding geometric content sometimes is facilitated by a perturbative expansion.

%%%%%%%%%%%%%%%%%%%%%%%%%%%%%%%%%%%%%%%%%%%%%%%%%%%%%%%%%%%%%%%%%%%%%%%%%%%%%%%%%%%%%%%%%%%%%%%%%%%%%%%%%%%%%%%%%%%%%%%%%%%%%%%

\section{Series, Nonlinear Geometric Objects, and Atlases}

Perturbative series expansions can also be useful for conceptual innovations.  For example, nonlinear realizations of the `group' of arbitrary coordinate transformations have tended to be invented with the help of a binomial series expansion for taking the symmetric square root of the metric tensor \cite{DeWittSpinor,OP,OPspinor}. The exponentiating technology of nonlinear group realizations \cite{IshamSalamStrathdee} is also at least implicitly perturbative.    While classical differential geometers defined nonlinear geometric objects (basically the same  as particle physicists' nonlinear group realizations as applied to coordinate transformations) \cite{Tashiro2,AczelGolab,SzybiakLie}, they generally provided no examples. 

Perhaps the most interesting example involves the square root of the (inverse) metric tensor, or rather a slight generalization for indefinite metrics.  The result is strictly a square root and strictly symmetric  using  $x^4 = ict$;  otherwise it is a generalized square root using the signature matrix  $\eta_{\alpha\beta}=diag(-1,1,1,1)$.
One has $ r^{\mu\alpha} \eta_{\alpha\beta} r^{\beta\nu} = g^{\mu\nu}$ and  $ r^{[\mu\nu]}=0. $
 Under coordinate transformations, the new components $ {r^{\mu\nu}}^{\prime}$  are nonlinear in the old ones \cite{OP,PittsSpinor}.  These entities augment tensor calculus and have covariant and Lie derivatives  \cite{Tashiro2,SzybiakCovariant}.

Defining the symmetric square root of a metric tensor might seem more of a curiosity for geometric completists than an important insight---but the symmetric square root of the metric makes an important conceptual difference with spinor fields used to represent fermions.  
Spinors in GR are widely believed to require  an orthonormal basis \cite{WeylElektronGravitation,CartanSpinor,SpinGeometry}. 
 But they don't, using $r^{\mu\nu}$ \cite{DeWittSpinor,OP,OPspinor,BilyalovSpinors}. One can have spinors in coordinates, but with metric-dependent transformations beyond 15-parameter conformal group \cite{OPspinor,IshamSalamStrathdee,BorisovOgievetskii,PittsSpinor}, the conformal Killing vectors for the unimodular metric density $\hat{g}_{\mu\nu} = (-g)^{-\frac{1}{4} } g_{\mu\nu}$. 
 Such spinors have Lie derivatives beyond conformal Killing vectors---often considered the frontier for Lie differentiation of spinors \cite[p. 101]{PenroseRindler2}---but they sprout new terms in $\pounds_{\xi} \hat{g}_{\mu\nu}$.  One can treat symmetries without surplus structure and an extra local $O(1,3)$ gauge group to gauge it away.  
  
%%%%%%%%%%%%%%%%%%%%%%%%%%%%%%%%%%%%%%%%%%%%%%%%%%%%%%%%%%%%%%%%%55

%\section{The Manifold Atlas and Nonlinear Geometric Objects}  

The (signature-generalized) square root of a metric, though not very familiar, fits fairly nicely into the realm of nonlinear geometric objects, yielding a set of components in every coordinate system (with a qualification) and a nonlinear transformation law.  The entity is useful especially if one wants to know what sort of space-time structure is necessary for having spin-$\frac{1}{2}$ particles in curved space-time \cite{WoodardSymmetricTetrad}.  Must one introduce an orthonormal basis, then discard much of it from physical reality by taking an equivalence class under local Lorentz transformations? Or can one get by without introducing anything beyond the metric and then throwing (most of?) it away?

A curious and little known feature of this generalized square root touches on an assumption usually made in passing in differential geometry.  Although one can (often) make a binomial series expansion in powers of the deviation of the metric from the signature matrix, and (more often) one can take a square root using generalized eigenvalues, there are exotic coordinate systems in which the generalized square root does not exist due to the indefinite signature \cite{BilyalovSpinors,PittsSpinor,DeffayetSymmetricTetrad}.  This fact is trivial to show in 2 space-time dimensions (signature matrix $diag(-1,1)$) using the quadratic formula: just look for complex eigenvalues.  The fact generally has not been noticed previously because most treatments (a great many are cited in \cite{PittsSpinor}) worked near the identity.  
Such a point could have been noticed some time ago by Hoek, but a fateful innocent inequality was imposed that restricted the coordinates (with signature $+---$).
\begin{quote}  ``We shall assume that [the metric tensor $g_{\mu\nu}$] is pointwise continuously connected with the Minkowski metric (in the space of four-metrics of Minkowski signature) and has $g_{00} > 0$.'' \cite{HoekDeservanNSymmetricGaugeGribov}
\end{quote}  
The lesson to learn is that there can be feedback from the fibers over space-time to the atlas of admissible coordinate systems for nonlinear geometric objects given an indefinite signature.  Naively assuming a maximal atlas causes interesting and quite robust entities not to exist.  Such a result sounds rather dramatic when expressed in modern vocabulary.  But coordinate inequalities are old  \cite{Hilbert2}, familiar \cite{MollerBook2}, and not very dramatic classically; coordinates can have qualitative physical meaning while lacking a quantitative one.  A principal square root is related to the avoidance of negative  eigenvalues of $g^{\mu\nu} \eta_{\nu\rho}$ \cite{HighamRoot87,HighamRoot}. Null coordinates are fine; the coordinate restriction is mild.
 Amusingly, coordinate \emph{order} can be important: if $(x,t,y,z)$ is  bad, switching to $(t,x,y,z)$ suffices \cite{BilyalovSpinors}.  
 
%%%%%%%%%%%%%%%%%%%%%%%%%%%%%%%%%%%%%%%%%

\section{Massive Gravity:  1965-72 Discovery of 2010 Pure Spin-$2$}

The recent (re)invention of pure spin-$2$ massive gravity \cite{deRhamGabadadze,HassanRosenNonlinear} used the symmetric square root of the metric, as did the first invention \cite{OP}, though not the second \cite{ZuminoDeser,MassiveGravity2}.  This problem has a curious history, from which \cite{OP} has been unjustly neglected.  That paper highly developed the symmetric square root of the metric perturbatively.  It derived a $2$-parameter family of massive gravities, which, I note, includes  two of the original three modern massive pure spin-$2$ gravities with a flat background metric. In light of the dependence of the space-time metric on the lapse function $N$ in a $3+1$ ADM split, there were only two Ogievetsky-Polubarinov theories  with any chance of being linear in the lapse (hence having pure spin-$2$ \cite{DeserMass}), though the naive cross-terms are rather discouraging.  These are the $n=\frac{1}{2}, $ $p=-2$ theory built around $ \delta^{\alpha}_{\mu} (g^{\mu\nu}\eta_{\nu\alpha} \sqrt{-g}^2)^{\frac{1}{2}}$, a theory reinvented as equation (3.4) of \cite{HassanRosenNonlinear}, and the $n=-\frac{1}{2},$ $p=0$ theory built around $ \delta_{\alpha}^{\mu} (g_{\mu\nu}\eta^{\nu\alpha} \sqrt{-g}^0)^{\frac{1}{2}}$. 
 A truly novel third theory is now known  \cite{HassanRosenNonlinear}.
A second novel modern result is the  nonlinear field redefinition of the shift vector \cite{HassanRosenSchmidtMay}, which allows the square root of the metric to be linear in the lapse. 

%While Ogievetsky and Polubarinov suggested that the pure spin-$2$ massive gravities were especially interesting (p. 195), they were not  concerned about the spin-$0$ ghost that most of their theories had.  Neither did they anticipate that linearly pure spin-$2$ theories generically were mixtures of spin-$2$ and spin-$0$ nonlinearly \cite{DeserMass}.  Only two are pure spin-$2$ nonlinearly, but $2$ is many more than $0$.  One can imagine much more rapid progress in the modern revival of massive gravity if (\cite{OP}) had been remembered; one could test its catalog of theories rather than starting over. 

More striking than the proposal of such theories long ago is the fact that in 1971-2 Maheshwari already showed that one of the Ogievetsky-Polubarinov theories had pure spin-$2$ \emph{nonlinearly} \cite{MaheshwariIdentity}!  Thus the Boulware-Deser-Tyutin-Fradkin ghost \cite{DeserMass,TyutinMass} (the negative energy sixth degree of freedom that is avoided by Fierz and Pauli to linear order but comes to life nonlinearly) was \emph{avoided before it was announced}.  Unfortunately Maheshwari's paper made no impact, being cited only by Maheshwari in the mid-1980s.  With Vainshtein's mechanism also suggested in 1972, \cite{Vainshtein}, there was no seemingly insoluble problem for massive spin-$2$ gravity in the literature.    Massive spin-$2$ gravity was largely ignored from 1972 until \emph{c.} 2000 largely because of failure to read Maheshwari's paper.  This example illustrates the point  \cite{ChangWater} that the history of a science has resources for current science.

%%%%%%%%%%%%%%%%%%%%%%%%%%%%%%%%%%%%%%%

\section{Conclusions }

The considerations above support the idea that progress in knowledge about gravity can be made by overcoming various barriers, whether between general relativity and particle physics, or between physics and the history and philosophy of science.   GR does not need to be treated \emph{a priori} as exceptional, either in justifying choosing GR over rivals or in interpreting it. GR is well motivated non-mysteriously using particle physicists' arguments about the exclusion of negative-energy degrees of freedom, arguments that leave only a few options possible.  To some degree the same holds even for the context of discovery of GR, given the renewed appreciation of Einstein's  ``physical strategy.'' 

Because conceptual problems of GR often can be resolved, there is no need to treat it as \emph{a priori} exceptional in matters of interpretation, either. 
Regarding gravitational radiation, Feynman reflected on the unhelpfulness of GR-exceptionalism:
\begin{quote}
What is the power radiated by such a wave?  There are a great many people who worry needlessly at this question, because of a perennial prejudice that gravitation is somehow mysterious and different---they feel that it might be that gravity waves carry no energy at all.  We can definitely show that they can indeed heat up a wall, so there is no question as to their energy content. \cite[pp. 219, 220]{Feynman}  
\end{quote}
 The conservation of energy and momentum---rather, energies and momenta---make sense in relation to Noether's theorems.  Change, even in local observables, is evident in the Hamiltonian formulation, just as in the Lagrangian/4-dimensional geometric form. 

To say that GR should not be treated as \emph{a priori} exceptional is not to endorse the strongest  readings of the claim that GR is just another field theory, taking  gauge-fixing and perturbative expansions as opening moves.  The mathematics of GR  logically entails some distinctiveness, such as the difference between external coordinate symmetries (with a transport term involving the derivative of the field) and internal symmetries as in electromagnetism and Yang-Mills.   
 Identifying such distinctiveness requires reflecting on the mathematics and its meaning, as well as gross features of embodied experience,  but it does not require conjectures  about the trajectory of historical progress or divination of  the spirit of GR.

Series expansions have their uses in GR.  Einstein's failure to think perturbatively in 1917 about the cosmological constant generated lasting confusion and surely helped to obscure massive spin-$2$ gravity as an option.  Many of the (re)inventions of the symmetric generalized square root of the metric began perturbatively. It permits spinors in coordinates,  a fundamental geometric result, just as was Weyl's 1929 impossibility claim. Perturbative methods should not always be used or always avoided; they are one tool in the tool box for the foundations of gravity and space-time.

%%%%%%%%%%%%%%%%%%%%%%%%%%%%%%%%%%%%%%%%%%%%%%%%%%%%

%\bibliography{Pitts}  
%\bibliographystyle{apalike} %{elsart-harv}  
%%

\end{document}